\documentclass{report}

\usepackage{amssymb}
\usepackage{amsmath}
\begin{document}
\begin{center}
\title*{\large\bf On the   integrable inhomogeneous Myrzakulov I equation}\\
\author*{\small Zhen-Huan Zhang$^a$, Ming Deng$^a$,
 Wei-Zhong Zhao$^{a}$\footnote{ E-mail
address: zhaowz100@163.com} and Ke Wu$^{a,b,}$ \footnote{
E-mail address: wuke@mail.cnu.edu.cn}\\
$^a$ Department of Mathematics, Capital Normal University, Beijing
100037, China \\
$^b$ KLMM, AMSS, Chinese Academia of Sciences, Beijing 100080,
China} \vskip 0.2cm
\end{center}
\begin{center}
\begin{minipage}{140mm}
\hspace{0.5cm} {\small {\bf Abstract}\\
 By using the prolongation
structure theory proposed by Morris, we give  a (2+1)-dimensional
integrable inhomogeneous Heisenberg Ferromagnet models, namely,
the inhomogeneous Myrzakulov I equation. Through the motion of
space curves endowed with an additional spatial variable, its
geometrical equivalent counterpart is also presented.
}
\end{minipage}
\end{center}
\section *{1. Introduction}
The Heisenberg Ferromagnet (HF) equation,
\begin{eqnarray}
\mathbf{S}_t&=&\mathbf{S}\times\mathbf{S}_{xx},\label{eq:HF}
\end{eqnarray}
which describes the motion of the magnetization vector of the
isotropic ferromagnets is an important integrable system. The
corresponding integrable inhomogeneous HF equation was derived by
Balakrishnan [1],
\begin{eqnarray}
{\bf S}_t&=&(f{\bf S}\times {\bf S}_{x})_x+h{\bf
S}_x,\label{eq:Bala}
\end{eqnarray}
where the linear functions $f$ and $h$ take $f=\mu_1x+\nu_1$ and
$h=\mu_2x+\nu_2$, and the coefficients $\mu_i$ and $\nu_i$,
$i=1,2$, are constants. On the basis of the prolongation structure
theory of Wahlquist and Estabrook [2], the integrable deformations
of the (inhomogeneous) HF model have been studied in Refs.[3,4].

The (2+1)-dimensional integrable extensions of a (1+1)-dimensional
integrable equation have been of interest. For the HF equation
(\ref{eq:HF}), several (2+1)-dimensional integrable extensions
have been constructed [5].  One of its important integrable
extensions is given by
\begin{eqnarray}
\mathbf{S}_t&=&(\mathbf{S}\times\mathbf{S}_y+u\mathbf{S})_x,\nonumber\\
u_x&=&-\mathbf{S}\cdot(\mathbf{S}_x\times\mathbf{S}_y),\label{eq:EHF}
\end{eqnarray}
 that is the Myrzakulov I (M-I) equation [5-7]. The M-I equation (\ref{eq:EHF}) is geometrical
and gauge equivalent to the well-known (2+1)-dimensional focusing
nonlinear Schr\"{o}dinger equation $NLS^+$ [6]
\begin{equation}
i\psi_t-\psi_{xy}-v\psi=0, \ \ \ \ v_x=2\partial_y|\psi|^2,
\label{eq:NLS}
\end{equation}
where $\psi$ is a complex function. Recently Morris's prolongation
structure theory for nonlinear evolution equation in two spatial
dimensions [8] has been successfully applied to analyze equation
(\ref{eq:EHF}) in Ref.[9].  It is noted that this prolongation
structure method is very simple and effective in the investigation
of M-I equation. For the inhomogeneous HF equation
(\ref{eq:Bala}), its (2+1)-dimensional integrable extensions have
not been studied so far. The purpose of this paper is to apply
Morris's prolongation structure method to construct the
(2+1)-dimensional integrable extension of equation
(\ref{eq:Bala}).

\section *{2. The  inhomogeneous integrable M-I equation}

 In order to investigate (2+1)-dimensional integrable extension of Eq.(\ref{eq:Bala}),  we introduce a new
vector $\mathbf{E}(\mathbf{S},\mathbf{S}_x,\mathbf{S}_y)$ and the
functions $f(x)$ and $g(x)$ into (\ref{eq:EHF}),
\begin{eqnarray}
\mathbf{S_t}&=&\{f\mathbf{S}\times\mathbf{S_y}+gu\mathbf{S}\}_x+\mathbf{E},\nonumber\\
&=&f\mathbf{S}\times\mathbf{S_{xy}}+f\mathbf{S}_x\times\mathbf{S}_y
+f_x\mathbf{S}\times\mathbf{S}_y+gu_x\mathbf{S}+gu\mathbf{S}_x+g_xu\mathbf{S}
+\mathbf{E}\nonumber\\
u_x&=&-\mathbf{S}\cdot(\mathbf{S_x}\times\mathbf{S_y}),\label{eq:EIHF}
\end{eqnarray}
where $\bf S\cdot\bf S=1$ and $\bf S\cdot E=0$. Multiplying Eq.
(\ref{eq:EIHF}) by $\mathbf{S}$ and using the restricting
relation,
$\mathbf{S}\cdot\mathbf{S_t}=\mathbf{S}\cdot\mathbf{S_x}=\mathbf{S}\cdot\mathbf{S_y}=0$,
we have $f=g$ and $f_x=g_x=0$. It implies that f and g should be
the constants.  Let us take $f=1$ and
\begin{eqnarray}
\mathbf{E}=\rho(x)\mathbf{S}_x+\nu(x)\mathbf{S}_y+\mu(x)\mathbf{S}\times\mathbf{S}_x,
\end{eqnarray}
where the functions $\rho(x)$, $\mu(x)$ and $\nu(x)$ need to be
determined. Thus equation (\ref{eq:EIHF}) can be rewritten as
\begin{eqnarray}
\mathbf{S_t}&=&\mathbf{S}\times\mathbf{S_{xy}}+\mathbf{S}_x\times\mathbf{S}_y
+u_x\mathbf{S}+u\mathbf{S}_x\nonumber\\
&+&\rho(x)\mathbf{S}_x+\nu(x)\mathbf{S}_y+\mu(x)\mathbf{S}\times\mathbf{S}_x
\nonumber\\
u_x&=&-\mathbf{S}\cdot(\mathbf{S_x}\times\mathbf{S_y}),\label{eq:EIHFE}
\end{eqnarray}

As done in Ref.[9], we first consider the prolongation structure
of (\ref{eq:EIHFE}) when ${\bf S}_t=0$. Setting
$\mathbf{W}=\mathbf{S_x}$, $\mathbf{T}=\mathbf{S_y}$ and taking
$\mathbf{S}$,$\mathbf{T}$,$\mathbf{W}$, and u as the new
independent variables, we can define the following set of two
forms,
\begin{eqnarray}
\alpha_a&=&dS_a\wedge dx-T_ady\wedge dx,\nonumber\\
\alpha_{a+3}&=&dS_a\wedge dy-W_adx\wedge dy,\nonumber\\
\alpha_{a+6}&=&(\mathbf{W}\times\mathbf{T})_adx\wedge dy+(\mathbf
{S}\times d\mathbf{T})_a\wedge dy+S_adu\wedge dy \nonumber\\
&+&uW_adx\wedge dy+(\rho W_a+\nu
T_a+\mu(\mathbf{S}\times\mathbf{W})_a)dx\wedge dy,\nonumber\\
\alpha_{10}&=&du\wedge
dy+\mathbf{S}\cdot(\mathbf{W}\times\mathbf{T})dx\wedge dy,\nonumber\\
\alpha_{a+10}&=&dT_a\wedge dy+dW_a\wedge dx,\nonumber\\
\alpha_{14}&=&(\mathbf{T}\cdot\mathbf{W})dx\wedge dy+S_a\cdot
dT_a\wedge dy,
\end{eqnarray}
where $a=1,2,3$, such that they constitute a closed ideal
$I=\{\alpha_i, i=1,2,\cdots,14\}$. Then we extend the above ideal
I by adding to it a set of one forms,
\begin{equation}
\Omega^k=d\xi^k+F^k(x, y, \mathbf{S},\mathbf{T},\mathbf{W},u)\xi^k
dx+G^k(x, y, \mathbf{S},\mathbf{T},\mathbf{W},u)\xi^k dy,\ \ \ \
k=1,2,\cdots ,n, \label{eq:omega}
\end{equation}
where $\xi^k$ is prolongation variable. In terms of the
prolongation condition, $d\Omega^k\subset\{I, \Omega^k\}$, we
obtain the following set of partial differential equations for
$F^k$ and $G^k$,
\begin{eqnarray}
& &\frac{\partial F^k}{\partial T_a}=\frac{\partial F^k}{\partial
u}=0, \ \ \ \ \frac{\partial
G^k}{\partial W_a}=0,\nonumber\\
&-&\frac{\partial F^k}{\partial S_a}T_a+\frac{\partial
G^k}{\partial S_a}W_a-\frac{\partial G^k}{\partial
u}\mathbf{S}\cdot(\mathbf{W}\times\mathbf{T})+(\frac{\partial
G^k}{\partial T_a}-\frac{\partial F^k}{\partial W_a})
\Big\{\big[\mathbf{S}\times(\mathbf{W}\times\mathbf{T})\big]_a\nonumber\\
&-&S_a(\mathbf{T}\cdot\mathbf{W})+u(\mathbf{S}\times\mathbf{W})_a
+ (\rho \mathbf{S}\times \mathbf{W}+\nu\mathbf{S}\times\mathbf{T}
+\mu\mathbf{S}\times(\mathbf{S}\times\mathbf{W}))_a {}\Big\}\nonumber\\
&-&[F,G]^k+\frac{\partial G^k}{\partial x}-\frac{\partial
F^k}{\partial y}=0,\label{eq:pde1}
\end{eqnarray}
where $[F,G]^k\equiv\sum_{l=1}^{n}F^l\frac{\partial G^k}{\partial
y^l}-\sum_{l=1}^{n}G^l\frac{\partial F^k}{\partial y^l}$. By
solving (\ref{eq:pde1}), we have the following solution,
\begin{equation}
F=\lambda\sum_{i=1}^{3}S_iX_i,\ \ \ \
G=(\rho+u)\sum_{i=1}^{3}S_iX_i+\sum_{i=1}^{3}(\mathbf{S}\times\mathbf{T})_iX_i,\
\ \ \ \ \mu=\nu=0, \label{eq:SFG}
\end{equation}
where  $X_i$, $i=1,2,3$, depend only on the prolongation variables
$\xi^k$ and have the commutation relation of the $su(2)$ Lie
algebra.

Let us turn now to define a set of 3-form $\overline{\alpha}_i$ as
follows,
\begin{eqnarray}
\overline{\alpha}_a&=&dS_a\wedge dx\wedge dt-T_ady\wedge dx\wedge
dt,\nonumber\\
\overline{\alpha}_{a+3}&=&dS_a\wedge dy\wedge dt-W_adx\wedge
dy\wedge dt, \nonumber\\
\overline{\alpha}_{a+6}&=&(\mathbf{W}\times\mathbf{T})_adx\wedge
dy\wedge dt+(\mathbf {S}\times d\mathbf {T})_a\wedge dy\wedge
dt+S_adu\wedge dy\wedge dt{}
\nonumber\\
& & {}+uW_adx\wedge dy\wedge dt-dS_a\wedge dx\wedge dy +(\rho
W_a+\nu T_a+\mu(\mathbf{S}\times\mathbf{W})_a)dx\wedge
dy\wedge dt,\nonumber\\
\overline{\alpha}_{10}&=&du\wedge dy\wedge
dt+\mathbf{S}\cdot(\mathbf{W}\times\mathbf{T})dx\wedge dy\wedge
dt,\nonumber\\
\overline{\alpha}_{a+10}&=&dT_a\wedge dy\wedge dt+dW_a\wedge dx\wedge dt,\nonumber\\
\overline{\alpha}_{14}&=&(\mathbf{T}\cdot\mathbf{W})dx\wedge
dy\wedge dt+S_a\cdot dT_a\wedge dy\wedge dt,
\end{eqnarray}
where $a=1,2,3$, such that they constitute a closed ideal. When
these two forms are null, we recover (\ref{eq:EIHFE}). Then we
introduce the following two forms,
\begin{equation}
\overline\Omega^k=\Omega^k\wedge dt+H^k_j\xi^j dx\wedge
dy+(A^k_jdx+B^k_jdy)\wedge d\xi^j, \ \ \ \ k=1,2,\cdots ,n,
\label{eq:somega}
\end{equation}
where the matrices of A and B  depend on the variables (x, y, t)
and the form of $\Omega^k$ is given by (\ref{eq:omega}), in which
$\lambda$ depends on the variables (x, y, t).  It is easily shown
that
\begin{eqnarray}
d\overline\Omega^k&=&\sum_{i=1}^{14}g^{ki}\overline{\alpha}_i+\sum_{j=1}^{n}\zeta_j^k\wedge\overline\Omega^j,
\end{eqnarray}
provided that the matrix H is given by
\begin{eqnarray}
H=GA-FB+A_y-B_x \label{eq:H}
\end{eqnarray}
and
\begin{eqnarray}
dH\wedge dx\wedge dy -\frac{\partial G}{\partial
T_a}(\mathbf{S}\times\mathbf{dS})_a\wedge dx\wedge dy -\lambda_y
S_aX_adx\wedge dy\wedge dt\nonumber\\
+\rho_xS_aX_adx\wedge dy\wedge dt -A_tGdx\wedge dy\wedge
dt+B_tFdx\wedge dy\wedge dt=0 .\label{eq:SH}
\end{eqnarray}
Substituting the expressions (\ref{eq:SFG}) of F and G  into
(\ref{eq:H}) and (\ref{eq:SH}), we obtain
\begin{eqnarray}
A=0, \ \ \ \  B=\frac{1}{\lambda}I,
\end{eqnarray}
and
\begin{eqnarray}
\lambda_t=-\lambda\lambda_y+\lambda\rho_x, \ \ \ \
\lambda_x=0.\label{eq:sp}
\end{eqnarray}
From Eq.(\ref{eq:sp}), we note that the function $\rho(x)$ take
the following expression,
\begin{eqnarray}
\rho(x)=\mu_3x+\nu_3,
\end{eqnarray}
where the coefficients $\mu_3$ and $\nu_3$ are constants. Thus the
 integrable inhomogeneous M-I equation is
\begin{eqnarray}
\mathbf{S_t}&=&\{\mathbf{S}\times\mathbf{S}_y+u\mathbf{S}\}_x
+(\mu_3x+\nu_3)\mathbf{S}_x
\nonumber\\
u_x&=&-\mathbf{S}\cdot(\mathbf{S_x}\times\mathbf{S_y}),\label{eq:IHF}
\end{eqnarray}
On imposing the reduction $\partial_y=\partial_x$, equation
(\ref{eq:IHF}) reduces to the (1+1)-dimensional inhomogeneous HF
equation (\ref{eq:Bala}) in which  $f=1$. By restricting
(\ref{eq:somega}) on the solution manifold, we obtain the Lax
representation of equation ({\ref{eq:IHF}})
\begin{eqnarray}
\xi_x&=&-F|_{X_i=-\frac{i}{2}\sigma_i}
\xi=\frac{i\lambda}{2}\sum_{i=1}^3 S_i\sigma_i\xi,\nonumber\\
\xi_t&=&-\frac{1}{B}\xi_y -\frac{1}{B}
G|_{X_i=-\frac{i}{2}\sigma_i}\xi \nonumber\\
&=&-\lambda\xi_y+\frac{i\lambda}{2}\sum_{i=1}^3[
(\rho+u)S_i\sigma_i+(\mathbf{S}\times\mathbf{T})_i\sigma_i]\xi.
\end{eqnarray}
where  $\sigma_i, i=1,2,3$, are Pauli matrices, and the spectral
parameter satisfies the nonlinear equation (\ref{eq:sp})

\section *{3. Geometrical equivalent counterpart}
By associating with the motion of Euclidean space curves endowed
with an extra spatial variable, Myrzakulov et al. showed that the
M-I  equation (\ref{eq:EHF}) is geometric equivalent to the
(2+1)-dimensional $NLS^+$ [6]. In order to give the geometrical
equivalent counterpart of (\ref{eq:IHF}), we fist give a brief
review on the motion of a curve. In general, the Serret-Frenet
equation associated with a curve is given by
\begin{eqnarray}
{\bf t}_s &=&k{\bf n}, \nonumber\\
{\bf b}_s &=&-\tau{\bf n}, \nonumber\\
{\bf n}_s &=&\tau{\bf b}-k{\bf t},\label{eq:SF}
\end{eqnarray}
where k and $\tau$ are the curvature and torsion of the curve,
respectively. This equation can also be rewrite as [9]
\begin{eqnarray}
{\bf N}_s=-\psi{\bf t},\ \ \ \  {\bf
t}_s=\frac{1}{2}(\psi^{\ast}\bf N+\psi\bf N^{\ast}),
\end{eqnarray}
where ${\bf N}=({\bf n}+i{\bf b})exp(i\int\limits_{-\infty}^s
ds'\tau)$, $\psi=k exp (i\int\limits_{-\infty}^s ds'\tau)$ and the
new frame ${\bf t}$,${\bf N}$ and ${\bf N}^{\ast}$ satisfies the
relations
\begin{eqnarray}
{\bf N}\cdot{\bf N}^{\ast}=2,\ \  {\bf N}\cdot{\bf t}={\bf
N}^{\ast}\cdot{\bf t}={\bf N}\cdot{\bf N}=0.\label{eq:RNT}
\end{eqnarray}
The temporal variation of ${\bf t}$,${\bf N}$ and ${\bf N}^{\ast}$
may be expressed as
\begin{eqnarray}
{\bf N}_t=\alpha{\bf N}+\beta{\bf N}^{\ast}+\gamma{\bf t}, \ \
{\bf t}_t=\lambda{\bf N}+\mu{\bf N}^{\ast}+\nu{\bf
t}.\label{eq:NT}
\end{eqnarray}
 Multiplying Eq.(\ref{eq:NT}) by $\bf N$ and $\bf
t$ and using the relation (\ref{eq:RNT}), we have
\begin{eqnarray}
{\bf N}_t=iR{\bf N}+\gamma{\bf t},\ \ {\bf
t}_t=-\frac{1}{2}(\gamma^{\ast}{\bf N}+\gamma{\bf
N}^{\ast}),\label{eq:NTT}
\end{eqnarray}
where $R(s,t)$ is real. Using the compatibility condition, ${\bf
N}_{ts}={\bf N}_{st}$, we get
\begin{eqnarray}
\psi_t+\gamma_s-iR\psi=0,\nonumber\\
R_s=\frac{1}{2}i(\gamma\psi^{\ast}-\gamma^{\ast}\psi).\label{eq:EE}
\end{eqnarray}
If the auxiliary function $\gamma$ and R can be expressed in terms
of $\psi$ and its spatial derivations, then equation (\ref{eq:EE})
will provide an evolution equation for the spatial and temporal
variation of the curvature and torsion of the curve as expressed
through $\psi$.

Let us identify $\mathbf S$ with the tangent of a Euclidean space
curve and endow the moving curve  with an additional spatial
variable y.   Thus equation (\ref{eq:IHF}) becomes
\begin{eqnarray}
\mathbf{t}_t&=&\mathbf{t}_x\times\mathbf{t}_y+\mathbf{t}\times\mathbf{t}_{yx}+u_x\mathbf{t}
+u\mathbf{t}_x +\rho\mathbf{t}_x \label{eq:TT},
\end{eqnarray}
where the subscript x denotes the arc length parameter.  On
imposing the compatibility condition,
$\mathbf{t}_{xy}=\mathbf{t}_{yx}$,
$\mathbf{n}_{xy}=\mathbf{n}_{yx}$ and
$\mathbf{b}_{xy}=\mathbf{b}_{yx}$, we obtain the y-part equation
of the tangent, normal and binormal vectors
\begin{eqnarray}
\mathbf{t}_y&=&-\frac{u_x}{\kappa}\mathbf{b}+\partial_x^{-1}(\kappa_y-\frac{\tau
u_x}{\kappa})\mathbf{n},\nonumber\\
\mathbf{n}_y&=&(u+\partial_x^{-1}\tau_y)\mathbf{b}-\partial_x^{-1}(\kappa_y-\frac{\tau
u_x}{\kappa})\mathbf{t},\nonumber\\
\mathbf{b}_y&=&-(u+\partial_x^{-1}\tau_y)\mathbf{n}+\frac{u_x}{\kappa}\mathbf{t},\label{eq:TY}
\end{eqnarray}
Substituting  (\ref{eq:TY}) and (\ref{eq:SF}) into (\ref{eq:TT}),
we have
\begin{eqnarray}
\mathbf{t}_t &=&\frac{1}{2}\big[(-i\psi_y+\rho\psi)\mathbf{N}
+(i\psi_y^*+\rho\psi^*)\mathbf{N^*}\big].\label{eq:STT}
\end{eqnarray}
Comparing  (\ref{eq:STT}) with (\ref{eq:NTT}), we get
\begin{equation}
\gamma=-(\rho\psi^*+i\psi^*_y),
\end{equation}
then substituting $\gamma$ into  (\ref{eq:EE}), we obtain the
following integrable evolution equation
\begin{equation}
i\psi_t-\psi_{xy}-i(\rho\psi)_x-R\psi=0,\ \ \ \
R_x=\frac{1}{2}\partial_y|\psi|^2 \label{eq:INLS}.
\end{equation}
It is the inhomogeneous extension of (\ref{eq:NLS}). The Lax
representation of  (\ref{eq:INLS}) is given by
\begin{eqnarray}
\Phi_x=U\Phi,\ \ \ \
\Phi_t=V\Phi+\lambda\Phi_y,
\end{eqnarray}
in which
\begin{equation}
U=\left(\begin{array}{ll}i\lambda/2&\psi/2\\
-\psi^{\ast}/2&-i\lambda/2\end{array}\right),\ \ \ \
V=\left(\begin{array}{ll}
-\frac{i}{2}R+\frac{i}{2}\lambda\rho&-\frac{i}{2}\psi_y+\frac{1}{2}\rho\psi\\
-\frac{i}{2}\psi^{\ast}_y-\frac{1}{2}\rho\psi^{\ast}&\frac{i}{2}R-\frac{i}{2}
\lambda\rho\end{array}\right),
\end{equation}
and the spectral parameter ¦Ë satisfies  Eq.(\ref{eq:sp}).
\section *{4. Summary}

We have investigated a possible  integrable inhomogeneous M-I
equation by using Morris's prolongation structure theory. Under
the reduction $\partial_y=\partial_x$, the (2+1)-dimensional
integrable equation which we obtained in this paper, i.e.,
Eq.(\ref{eq:IHF}), reduces to a special case of inhomogeneous HF
equation (\ref{eq:Bala}). It has been noted that there exist
several (2+1)-dimensional integrable extensions for the HF
equation (\ref{eq:HF}), such as M-VIII, Ishimori and M-IX
equations [5-7, 11]. Therefore,  whether there are more general
(2+1)-dimensional integrable inhomogeneous extensions of
(\ref{eq:EHF}) is a question for the future.
\\
\\
\\
{\bf Acknowledgements}\\
We would like to express our thanks to Moningside Center, CAS,
part of the works was done when we were joining the workshop of
Math-Phys there. We are also very grateful to Prof. R. Myrzakulov
for his interest and helpful discussions. This work is partially
supported by German(DFG)-Chinese(NSFC) Exchange Programme
446CHV113/231, NKBRPC (2004CB318000), Beijing Jiao-Wei Key project
(KZ200310028010), NSF projects (10375038
and 90403018) and SRF for ROCS(222225), SEM.\\
\\
{\bf References}

[1] R. Balakrishnan, J. Phys. C 15 (1982) L1305.

[2] H.D. Wahlquist, F.B. Estabrook, J. Math. Phys. 16 (1975) 1.

[3] D.G. Zhang, G.X. Yang, J. Phys. A 23 (1990) 2133.

[4] W.Z. Zhao, Y.Q. Bai and K. Wu,  Phys. Lett. A 352 (2006) 64.

[5] R. Myrzakulov , G.N. Nugmanova and R.N. Syzdykova, J. Phys. A
31 (1998) 9535.

[6] R. Myrzakulov ,  S. Vijayalakshmi,  G.N. Nugmanova  and M.
Lakshmanan,

Phys. Lett. A 233 (1997) 391

[7] L. Martina, Kur. Myrzakul, R.Myrzakulov, G.Soliani, J. Math.
Phys.  42 (2001)  1397.

[8] H.C. Morris, J. Math. Phys.17 (1976) 1870.

[9] Y. Zhai, S. Albeverio, W.Z. Zhao and K. Wu, J. Phys. A 39
 (2006) 2117.

[10] G.L. Lamb, J. Math. Phys. 18 (1977) 1654.

[11] Y. Ishimori, Prog. Theor. Phys. 72 (1984) 33.
\end{document}